\documentstyle[12pt]{article}
\begin{document}
\begin{center}
{\Large\bf On Valence Gluons in Heavy Quarkonia}
\footnote{This work was supported by the RFBR grants No.96-02-18137 and
96-15-96423} \\[2mm]
\vspace*{.3cm}

S.B.Gerasimov\\[2mm]
\vspace*{.3cm}
Bogoliubov Laboratory of Theoretical Physics, JINR, Dubna \\[2mm]
\end{center}
\abstract
\noindent
To include the explicit valence gluon degrees of freedom
into spectroscopy of the lowest states  of heavy quarkonia,
we consider, within an adiabatic model, the properties of the lowest
hybrid $\bar{Q}Qg$-mesons and estimate the effects of their
mixing with low-lying vector $\bar{c}c$-charmonia.
The perspectives of
compatibility of the resulting picture with data are discussed.

\vspace*{.3cm}

The spectroscopy of systems constructed of a heavy quark $Q$ and a
heavy antiquark $\bar Q$ is known to be well described by the phenomenological
potential models\cite {1}, where free parameters in the potentials are
determined by fitting the calculated observables to the data.The gluon degrees
of freedom are assumed to be integrated out in these models.The validity of
this procedure is however questionable due to the presence of slow,
 large-scale nonperturbative vacuum fluctuations of the gluon field, as it
follows from the lattice QCD approaches.\\
On the other hand, recent
progress in understanding the production and decay processes of heavy
quarkonia\cite {2} is related to the idea of the valence gluon admixture
in heavy quarkonium state vectors and the presence of the colour-octet
$\bar{Q}Q$ - subsystem inside the colour-singlet bound state
\begin{equation}
|heavy\;\;meson\rangle = a_0|Q\bar Q\rangle + a_1|Q\bar Q g\rangle  + ...
\end{equation}
Therefore, those quarkonium states which are outside the potential regime
should not be used to fix parameters in the potential models, while for
the states containing the gluon admixture the application of the potential
models alone should leave a sufficient "room" for improvement assigned
to subsequent inclusion of the gluon degrees of freedom.\\
The key idea of our approach is nonperturbative mechanism of
higher Fock - components ( i.e. the state vectors with the constituent
gluons, Eq.(1) ) generation in heavy quarkonia via the mixing with low-lying
hybrid states. In the hybrid states, gluons are confined in a bound state by
the confining interaction with the colour-octet quark core, which is assumed
to be represented by an effective potential acting in three-body systems.
Turning to this particular picture of the heavy hybrid composition, we
notice existence of slow ($\bar{Q}Q$) and fast ($g$) sub-systems.Therefore,
 it is natural to proceed\cite {3} in the spirit of the Born-Oppenheimer, or
adiabatic approximation, {\it i.e.} to solve first a relativistic wave
equation for the gluon moving in a (presumably, confining) field of fixed
center, and then to make use of the found gluon energy  $\epsilon_g$ as a part
of the potential entering into the Schr\"odinger equation for the slow $\bar
QQ$ - sub-system
\begin{eqnarray} (\frac {{\vec p}^2}{m_Q} +\frac {1}{6}
\frac {\alpha_s}{R_Q} + V_{\bar QQ}(\varepsilon_g) - E_Q)\Psi(\vec R_Q) = 0
\end{eqnarray}
We note that the "Coulomb" potential is repulsive here because
the $\bar QQ$ -pair is in the colour octet state.
Our further main assumptions are as follows.We take the lowest magnetic (M1)-  and
electric (E1)- modes for the spin-orbital wave function of gluons
\begin{eqnarray}
\vec Y_{j,l,m}^{M1} = \vec Y_{j,j,m}(\theta,\varphi)\mid_{j=1}\\
\vec Y_{j,l,m}^{E1} =[ \sqrt {\frac {2j}{2j+1}}\vec Y_{j,j-1,m}(\theta,\varphi) + \sqrt {\frac {j}{2j+1}}\vec Y_{j,j+1,m}(\theta,\varphi)]_{j=1}
\noindent
\end{eqnarray}
where $\vec{Y}_{j,l,m}(\theta,\varphi)$ are the vector spherical harmonics.
This means that for the $J^{PC}=1^{--}$- hybrid mesons we are going to consider,
the orbital momentum $l$ of a $\bar QQ$-system should be $l=0(1)$ for the
$M1(E1)$ - gluon modes.The radial part of the gluon wave function is assumed
to obey the Klein-Gordon (KG) equation with an external field including the
(strong) Coulomb potential and the squared form of the linear confinement
potential, properly scaled against analogous potentials for colourless
$\bar QQ$ -states ( the scaling factor being the ratio of the corresponding
Casimir operators equal to $9/4$).For the assumed interaction between the two
colour-octet, point-like particles, the KG-equation would be of the form
\begin{eqnarray}
(\vec p_g^2 + {{V_s}^{g}}^2(r_g) + 2\varepsilon_g {V_v}^{g}(r_g) - {{V_v}^{g}}^2(r_g) - {\varepsilon_g}^2)\psi(\vec r_g) = 0
\end{eqnarray}
As far as the colour "charge" of the $\bar QQ$ - sub-system is spatially
distributed, we define the "form factor-modified" potentials through
the folding integral
\begin{eqnarray}
V_{s(v)}(\vec r_g) = \int V_{s(v)}(\mid \vec r_g - \vec r^\prime \mid)\rho(\vec r^\prime) d^3 r^\prime
\end{eqnarray}
where the density function $\rho(\vec r)$ is related to an (unknown) wave
function of heavy quarks.\\

Finally, our calculation scheme acquires the
variation form.We take, as a trial wave function of quarks, simple expressions
of the exponential form ( with the pre-exponential centrifugal or nodal
factors) which contain one variable parameter.  This parameter propagates to
the gluon energy, and then it appears again in the equation of motion for
quarks. The last step is the minimization of the Schr\"odinger energy
functional leading to the numerical value of this variable parameter and all
energies,hence, to the hybrid meson mass.Summing up, with the trial wave
functions of exponential form and on the basis of the adiabatic approximation,
the  masses of the vector hybrid states were estimated to be 4.02 (or 4.21)
GeV for charmed quarks with mass $m_c$=1.4 GeV and a valence gluon of the
M1(or E1)-type,while for the bottom quarks with mass $m_b$ =4.8 GeV the
corresponding masses are 10.65 (or 10.75) GeV. The mean values of $r_g$ and
$R_{\bar{Q}Q}$ characterizing spatial extension of hybrid wave functions are
equal to .45 fm and .4 fm  for the charmed states, and .47 fm and .3 fm  for
the b-flavored hybrids.The approximate independence of the characteristics of
light particle (i.e. the gluon) on masses of quarks is familiar
manifestation of the heavy quark symmetry.  The values of the "Coulomb"
constant and slope of the linear potential have been taken equal to
$\kappa$=$\frac {4}{3} \alpha_s$=.49$ and $a$ =.16$ $GeV^2$.  The obtained
values are rather close to estimates from different models\cite {4}. In
particular, they are very close to the string model elaborated in
\cite {5} where the gluon-quark interaction has acquired the form following
from our formula (6) if
\begin{eqnarray}
\rho(\vec{r})=1/2(\delta(\vec{r}-\vec{r}_{Q}) + \delta(\vec{r}-\vec{r}_{\bar{Q}}))
\nonumber
\end{eqnarray}
It is quite natural to expect that proper estimation  of the hybrid meson(s)
mixing with nearby quarkonia will be important to understand
some peculiarities
of the charmonium spectra and decays slightly over $4 GeV$\cite {6}.\\
As a first step, we confine
ourselves to consideration of the four-level mixing in the charmonium
spectrum choosing the ground state $J/ \Psi(1S)$, $\Psi(2S)$,
$\Psi(3S)$ and the
presumed hybrid state $H_{c}$ with the calculated mass around $4 GeV$ as
mixing states.  The nondiagonal elements $m_{n H}$ in the 4x4 - mass-matrix
\begin{eqnarray}
& &m_{n H}=\langle H_c; Q\bar Q g|{\cal{H}}_{int}|n S; Q\bar Q\rangle\\
{\cal{H}}_{int}&=& g_s\sum_i \frac{1}{2}\vec{\lambda}(i)(i\epsilon_g)(\vec{A}^{E1}(r,\Theta,\phi)\cdot \vec{r}(i))\delta(\vec{r}-\vec{r}(i))+\nonumber\\
&+&\frac{g_s}{2m_Q}\sum_i \frac{1}{2}\vec{\lambda}(i)(\sigma (i)\cdot\vec{A}^{M1}(r,\Theta,\phi))\delta(\vec{r}-\vec{r}(i))\nonumber
\end{eqnarray}
are calculated with the explicit radial wave functions
\begin{equation}
R_{3S}(r)= N_{3}\cdot (1- a_{3}(\gamma r)^m +b_{3}(\gamma r)^{2m})\exp{(1/2(\gamma r)^m)}, m=\frac{4}{3}
\end{equation}
\begin{eqnarray}
R_{1S}=R_{3S}("3"\rightarrow"1";a=b=0), \\
R_{2S}=R_{3S}("3"\rightarrow"2";b=0),
\end{eqnarray}
parametrized to reproduce
approximately the spatial dimensions ( {\it i.e.}$<r^2>$),the location of the
radial function nodes and the values of the wave functions of
the $1S$- ${3S}$ -quarkonia states at "zero" distance, which correspond
to the QCD-motivited potentials (e.g.\cite {7} and references therein).\\
As representative sets of $[\gamma_n;a_n;b_n]$ for the $(nS)$-states of
charmonia we take $[\gamma_1=.883;a_1=b_1=0]$, $[\gamma_2=.715;a_2=.57;b_2=0]$
$[\gamma_3=.628;a_3=1.101;b_3=.197]$, where all $\gamma$'s are in units of
$GeV$. The lowest vector hybrid state $h_c(g_{M1}Q\bar Q)$ with
the $M1$-type gluon mode, {\it i.e.}, with $l_g=1$ and $L_{Q\bar Q}=0$
also called "the gluon-excited state", should presumably be rather
narrow due to the dynamical selection rule discussed in a number of earlier
papers\cite {8,9}, which prevents the decays of this state into the
ground-state charmed mesons.
This selection rule is not acting for hybrids with the $E1$-type gluon mode
$(l_g=0, L_{Q \bar Q}=1)$, or "the quark-excited state", which should
therefore have very large width\cite {10}. Hence, in what follows, we
consider the mixing of the low-lying vector charmonia with the gluon-excited,
$M1$-type vector hybrid state. The radial wave function of this hybrid meson
is taken in the factorized form in accord with the adopted adiabatic
approximation
\begin{eqnarray}
R_{gQ \bar Q}(r_g,R_Q)= N_{g}N_{Q} r_g\exp{(1/2(\alpha_{g} r)^m + 1/2(\beta_{Q} R_{Q})^m)}, m=\frac{4}{3}
\end{eqnarray}
with the numerical values $\alpha_g=1.235$ GeV, $\beta_Q=.973$ GeV.
The nondiagonal elements of the symmetric $4 \times 4$- mass-matrix
have been calculated as matrix elements of the interaction
hamiltonian (7) over the
$nS$-charmonia states $(n=1,2,3)$, and the ("fourth") hybrid state and
their values are: $m_{n4}=.25, .074, .044 \; GeV$ for $n=1, 2, 3$,
respectively, with all other nondiagonal elements equal to zero. \\
To obtain
"physical" eigenvalues of the diagonalized matrix near to the known
masses of the $\Psi$-family, we take the "bare" masses, which stand along the
main diagonal having values $m_{nn}=3.153; 3.695; 4.05 \; GeV$ for
$n=1,2,3$ and $m_{44}(h_c)=4.07 GeV$. The diagonalization procedure leads to
the "physical" masses:  $m(J/\Psi)=3.089 [3.097], m(\Psi(2S))=3.685
[3.686], m(\Psi(3S))=4.03 [4.04], m(H_c)=4.17 [4.16]$ where  masses of the
known $\Psi$-mesons in $GeV$ are indicated in parentheses.  The corresponding
eigenfunctions reveal the following quark-gluon configuration mixing
\begin{eqnarray}
J/{ \Psi} = .968 \psi(1S)+.0302 \psi(2S)+.0113 \psi(3S)-.247 \psi(h_c),\\
\Psi(2S) = -.0628 \psi(1S)+.989 \psi(2S)+.0161 \psi(3S)-.134 \psi(h_c),\\
\Psi(3S) = -.0908 \psi(1S)-.0697 \psi(2S)+.940 \psi(3S)-.321 \psi(h_c), \\
H_c = .223 \psi(1S)+.141 \psi(2S)+.332 \psi(3S)+.906 \psi(h_c) .
\end{eqnarray}
Assuming the dynamical dominance of the quarkonia-components in the mentioned
states while computing the leptonic decay widths of the corresponding
vector mesons, we can compare the model and phenomenological ratios
of the meson wave functions "at the zero relative $Q -\bar Q$-distance":
\begin{equation}
 R_{J/{\Psi}}^2(0) : R_{2S}^2(0): R_{3S}^2(0) : R_{H_c}^2(0) =
 1 : .65 [.64] : .34 [.27 \pm .05] : .34 [.29 \pm .09]  ,
\end{equation}
where the values in parentheses are calculated using the proportionality
between $R_{V}^2(0)$ and $m^2(V)\Gamma(V \rightarrow l^{+}l^{-})$.
The puzzling equality of the leptonic widths of the $\Psi(4.04)$ and
$\Psi(4.16)$ states is explained in our approach by a coherent sum of the
admixture (separately, looking small) amplitudes of the $(1S)-(3S)$ quarkonia
states in the dominantly hybrid $\Psi(4.16)$-resonance.Whether our
interpretation of this phenomenon can be experimentally distinguished from the
Ono-Close-Page scenario\cite {6,11} (the approximately equal partition of
the $(3S)$-state and the $H_c$ -state between the $\Psi(4.04)$ and
$\Psi(4.16)$ charmonium state vectors) remains to be considered. To deal with
many other interesting implications of the valence gluon admixtures in the
low-lying charmonia, the involvement problem of the broad $(g_{E1} Q\bar
Q)$-type hybrid state(s) needs to be clarified.\\
\vspace*{.3cm}
This work was supported in part by the Russian Foundation for Basic Research
grants 96-02-18137 and 96-15-96423.\\

\end{document}